TITLE: The Impact of Simple, Brief, and Adaptive Instructions within Virtual Reality Training: Components of Cognitive Load Theory in an Assembly Task

AUTHORS:
Rebecca L. Pharmer[1], Christopher D. Wickens[1], Lucas Plabst[1,] Benjamin A. Clegg[2], Leanne M. Hirshfield[3], Joanna E. Lewis[4] , Jalynn B. Nicoly [3] , Cara A. Spencer [3], and Francisco R. Ortega[1]

[1]Colorado State University, Fort Collins, CO
[2]Montanna State University, Bozeman, MT
[3]University of Colorado Boulder, Boulder, CO
[4]University of Northern Colorado, Greely, CO

ACKNOWLEDGEMENTS: This research was supported by the Office of Naval Research conducted under grant number N00014-23-1-2298.

# Abstract

**Objective.** The study examined the effects of varying all three core elements of cognitive load on learning efficiency during a shape assembly task in virtual reality (VR).

**Background.** Adaptive training systems aim to improve learning efficiency and retention by dynamically adjusting difficulty. However, design choices can impact the cognitive workload imposed on the learner. The present experiments examined how aspects of cognitive load impact training outcomes.

**Method.** Participants learned step-by-step shape assembly in a VR environment. Cognitive load was manipulated across three dimensions: Intrinsic Load (shape complexity), Extraneous Load (instruction verbosity), and Germane Load (adaptive vs. fixed training). In adaptive training (experiment 1), difficulty increased based on individual performance. In fixed training (experiment 2), difficulty followed a preset schedule from a yoked participant.

**Results.** Higher Intrinsic Load significantly increased training times and subjective workload but did not affect retention test accuracy. Extraneous Load modestly impacted training time, with little impact on workload or retention. Adaptive training shortened overall training time without increasing workload or impairing retention. No interactions were observed between the three types of load.

**Conclusion.** Both Intrinsic and Extraneous Load increased training time, but adaptive training improved efficiency without harming retention. The lack of interaction between the elements suggests training benefits can be worth seeking within any of the components of cognitive load.

**Application:** These findings support the use of VR adaptive systems in domains such as manufacturing and military service, where efficient assembly skill acquisition is critical. Tailoring difficulty in real-time can optimize efficiency without compromising learning.

# The Impact of Simple, Brief, and Adaptive Instructions within Virtual Reality Training: Components of Cognitive Load Theory in an Assembly Task

The goal of improving training efficiency while maximizing learning retention is present across multiple industries and domains. Efficient training ensures that learners not only acquire skills faster but also retain and apply them effectively. In fields like surgery, aviation, or technical operations, where precision and accuracy are crucial, well-designed training can significantly impact performance and reduce safety-critical errors. As technology has advanced, training systems have addressed this need through algorithms that modulate instruction in real-time based on input from a human learner. These adaptive training systems are automated training programs that are intended to perform tutoring functions, like those traditionally carried out by a human tutor (Van Lehn, 2011). The effectiveness of these systems has been attributed to their ability to adapt to effectively accommodate individual differences in students. This approach avoids overloading the mental resources of the learner and instead seeks to provide an optimal level of difficulty in training to promote better retention of materials (Park & Lee, 2004; Durlach & Ray, 2011; Durlach, 2019).

Unfortunately, designing training systems that both facilitate quicker learning and ensure an appropriate level of competency upon completion is no easy feat. Indeed, Durlach (2019) discusses the balancing act that training system designers face between achieving optimum system usability, implementation of empirically supported pedagogical practices, and managing the cognitive load imposed on the learner. By applying theories of attention and learning to system design, the current pair of experiments aim to investigate how cognitive load manifests within adaptive training systems and whether that load can be assessed through behavioral outcomes like learning retention and training efficiency. We further investigate whether self-report measures of cognitive workload reflect the amount of load imposed on the learner. To accomplish this, we specifically pull from guidance set by Kalyuga and Sweller's (2004) Cognitive Load Theory.

**Cognitive Load Theory (CLT).** Cognitive Load Theory is a well validated theory of instruction in which the load imposed upon the learner in an instructional setting is assumed to come from different sources (Chandler & Sweller, 1991; Kalyuga & Sweller, 2004). The first source, *Intrinsic Load*, is the mental workload imposed by the inherent properties of the task to be learned. Some are more complex than others. For example, the task of flying an airplane has a greater Intrinsic Load than of driving a car in that the former contains six interacting axes of control, while the latter has two. Intrinsic Load has many causes; among the most salient are the working memory demands of the task and the number of interacting elements required to learn the task. The second load source, *Extraneous Load*, results from unnecessary sources of distraction and workload in the learning environment. For example, causes of high Extraneous Load are overly wordy instructions, difficult-to-read text, or the wide spatial separation between two pieces of information necessary for learning (e.g., the need for the learner to look up unfamiliar material contained in instructions in a separate handbook; Mayer & Marino, 2003), or a poor computer interface design. The third and final source is referred to as *Germane Load,* which is defined as the resources productively invested in the learning or mastery of the material. For example, it is well established that actively rehearsing task material, rather than just passively viewing (or hearing) it, while increasing cognitive load, produces better learning, and hence better retention of the material after training (Kalyuga & Sweller, 2004). A similar benefit is conferred on self-testing as well as notetaking (Roediger & Karpicke, 2006; Dunlosky et al., 2013), despite the added cognitive load that these activities impose.

The overall message of CLT is that sources of Extraneous Load should be eliminated, and the Intrinsic Load should be maintained sufficiently moderate during training so that there are ample cognitive resources available to the learner to allocate to the sources of Germane Load (Kalyuga & Sweller, 2004; Van Merriënboer, et al, 2006; Sweller, 2011). While an extensive body of literature has examined and validated each of these load effects (e.g., Sweller, 2011, de Jong, 2018, Mayer et al., 2008), there are certain fundamental elements that appear to be missing or inadequately represented in this collective knowledge. First, a vast majority of the empirical experimental validation of CLT has focused on the teaching of cognitive knowledge of a relatively abstract or cognitive nature, such as problem solving (Sweller, 1976; Schnotz & Kürschner, 2007). Less examined is the relevance of the three dimensions of cognitive load to the acquisition of hands-on perceptual-motor skills, particularly those with a perceptual-motor and procedural component, typical of the assembly of equipment, or complex checklist or start-up-procedures. It is our intention to examine CLT in the specific context of such a procedural skill. Additionally, while many studies have experimentally varied and validated the effects of one or two of the CLT variables in combination, few have varied all three, or even two in an orthogonal experimental design (but see Zu et al., 2019), as we do in the current research. In particular, there has been insufficient efforts to examine the influence of Germane Load, perhaps because of the added complexity of its definition as described below. Most relevant here appears to be the large body of work in classroom education in which certain learning strategies, such as self-testing or note taking, that are obvious sources of increased cognitive effort, have been well validated to improve the long-term acquisition of knowledge (see Dunlosky et al., 2013, Soderstom & Bjork, 2015, and Rhodes, Cleary & DeLosh., 2020 for reviews).

Related to the ideas regarding Germane Load, the cognitive construct of *desirable difficulties* of the material to be mastered has also emerged (Druckman & Bjork, 1994). Mapping on to some similar concepts to those found in CLT, this work suggests that the right kind of challenges within learning enhance outcomes, yet explicitly states that not all difficulties are "desirable" (e.g., Bjork & Bjork, 2020). The level of difficulty needs to be grounded in the learner's current knowledge. Hence, for a novice it might be better to start learning with a simple version (low Intrinsic Load) which supplies sufficient mental difficulty at that stage, but there will be a need to move beyond that simplicity and increase the demands across training in order to still challenge the learner later in learning. This approach can be referred to as "increased difficulty training" and it involves imposing progressively increasing levels of cognitive load. A meta-analysis by Carolan et al. (2013) validated the success of increasing difficulty training, although only when that increase was adaptively administrated, a qualification important in the current experiment. One key implication of desirable difficulties and Germane Load is the notion that increasing load during training will not necessarily impair learning, so long as that higher load is allocated to beneficial mental activities that translate to superior training outcomes (Sweller, 1976; Bjork & Bjork, 2020).

The challenge of manipulating Germane Load arises from the two different uses of the term "resources". On the one hand, the resource-demand of any task in a learning environment can be well defined by objective metrics such as those applied to Intrinsic Load addressing task complexity (Boag et al., 2006) working memory load (Hart & Wickens, 2010; Hirshfield et al., 2023), or those applied to Extraneous Load, such as the objective listing of load factors unrelated to the task (e.g., text legibility, visual angle separation between information sources). It is considerably more challenging to objectively quantify the resource demand of a Germane Load strategy like notetaking, self-testing, or rehearsing. On the other hand, it is only the actual resources invested by the learner while learning a task that can truly define their experienced Germane Load. These three factors require a careful triangulation to quantify Germane Load: 1) A learning strategy must be applied (or mandated to the learner in an experimental environment, for example, mandate that the learner engage in self-testing); 2) The strategy must generate an actual increase in resources invested; and 3) The strategy must have been demonstrated to be effective in actual learning. The third factor needs to be indexed by performance in some delayed retention or transfer task (i.e., performance on the target task some substantial time after training has been completed). This is because performance during acquisition is not an adequate basis for inferences about the quality of learning that has taken place (Schmidt & Bjork, 1992; Soderstrom & Bjork, 2015). The combination of all three of these factors is frequently missing from experiments that have tried to explicitly assess Germane Load, particularly with the training of a procedural skill. In the set of experiments described here, we attempt to address all three of the above factors in manipulating Germane Load, as we couple this manipulation with orthogonal manipulations of both Intrinsic and Extraneous Load while training a procedural skill.

Specifically, we hypothesize that:

- **H1**: Increasing *Intrinsic Load* will decrease training efficiency, retention test accuracy, and the mental load imposed on the learner.
- **H2**: Increasing *Extraneous Load* will show the same effects as in H1.

- **H3**: Optimizing *Germane Load* through adaptive training (experiment 1) will reduce time spent training and improve retention test accuracy but will lead to higher overall perceived workload from the imposed "desirable difficulties", as compared to fixed training (experiment 2).

## Experiments

To investigate our three hypotheses, we designed a 2 × 2 × 2 experimental design that addresses all three of the above factors in manipulating Germane Load, as we couple our Germane Load manipulation with orthogonal manipulations of both Intrinsic and Extraneous Load, while training a procedural skill. Figure 1 provides an overview of this experimental design.

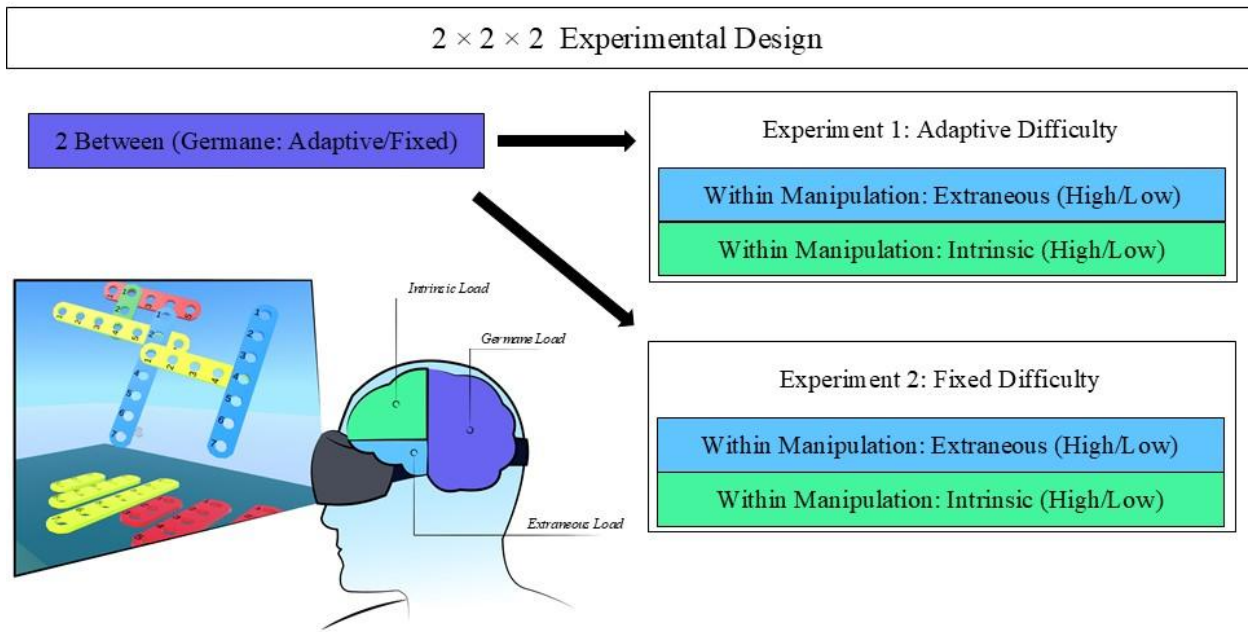

*Figure 1*. Our novel 2 × 2 × 2 experimental design to manipulate Germane, Intrinsic, and Extraneous Load.

As shown in Figure 1, our 2 × 2 × 2 design enables us to manipulate all three aspects of CLT as we couple our manipulation of Germane Load with orthogonal manipulations of both Intrinsic and Extraneous Load while training a procedural skill. These studies are further described next, but we first detail the assembly task that all participants undertook.

**Training Environment: Shape Assembly in Virtual Reality**

All studies took place in VR where participants were asked to assemble shapes by following a set of instructions. For the assembly task, participants were instructed to learn the assembly of shapes constructed out of bars of varying lengths and colors. Figure 2 provides an example of the shape stimuli.

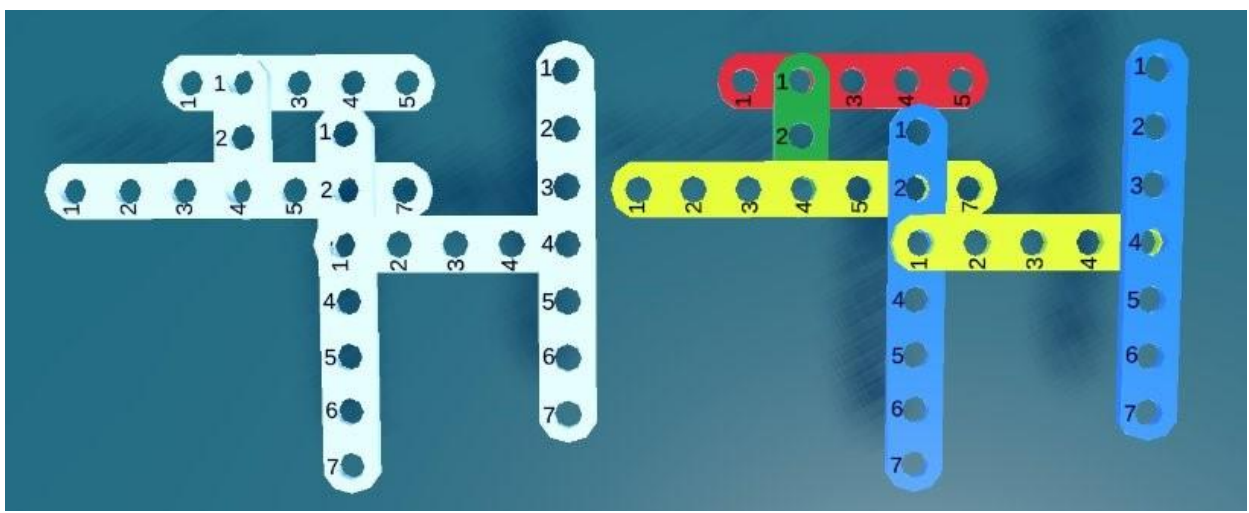

*Figure 2.* Example shape stimuli. The white shape on the left represents those in the Low-Intrinsic Load conditions and the colorful shape on the right represents those in the High-Intrinsic Load conditions.

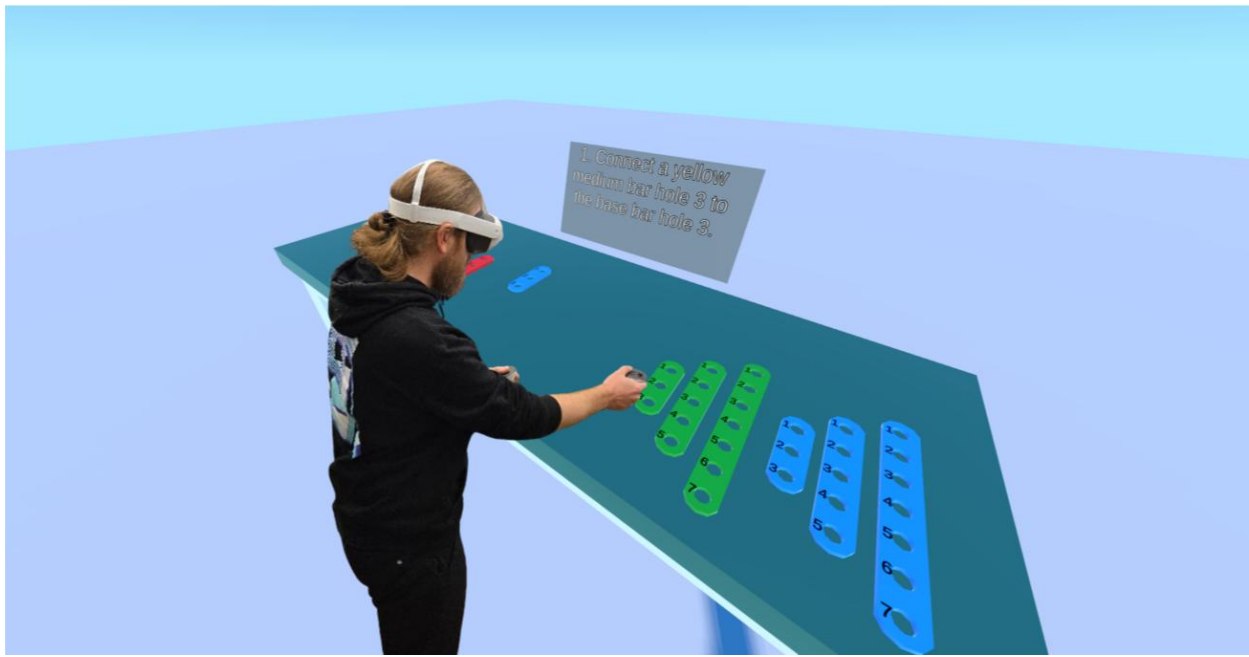

*Figure 3.* A user in VR assembling a shape.

*Experimental Platform.* This study environment was implemented using Unity Engine 2022.3.7f1 (Unity Technologies, 2022) with the XR Interaction Toolkit 0.9.4. (Unity Technologies, n.d.). In the experiment, a Meta Quest 3 (Meta, n.d.) VR head-mounted display (HMD) was used. The application ran on the HMD itself with a consistent 120 frames per second. This target was identified by Wang et al. (2023) to be an optimal frame rate for VR, as it increases user performance and reduces the chance of simulation sickness significantly. Upon putting on the HMD, participants were instructed to calibrate the interpupillary distance, ensuring optimal display clarity.

**Experimental Manipulations of Germane, Intrinsic, and Extraneous Load.**

**Adaptive training and desirable difficulties.** As we have described above, we have chosen an increasing difficulty strategy as a means of imposing Germane Load. In our experiment we do not manipulate Germane Load to be either low or high as is typically done with Intrinsic and Extraneous Load, but rather to be optimal versus non-optimal. Optimal Germane Load is that which is imposed adaptively as a function of performance as training proceeds (Landsberg et. al., 2012); and we do so here by increasing, adaptively, the cognitive demands across training trials. We accomplish this by adaptively removing scaffolding supports within the nature of instruction, thereby forcing increasing reliance upon memory rather than visual information to perform the task, an increase we deem productive to learning (i.e., Germane). This increase will produce the "desirable difficulty" of the task for learning (Bjork & Bjork, 2020). In turn, we achieve this increase adaptively, rather than on a fixed schedule because only adaptive difficulty increases have been found to be productive in improving transfer (Carolan et al., 2013) and because adapting these increases to each learner's skill level is assumed to be more optimal than the "one size fits all" approach of increasing difficulty at the same fixed rate for every learner.

**Extraneous Load and Mixed Reality.** The paradigm we employ is intended to generalize to a learner, equipped with an augmented reality HMD, using information provided on that display to guide them through a sequence of assembly steps on a physical piece of equipment. The choice of an HMD in this context is based very directly upon the reduction of Extraneous Load. This is because in conventional assembly, the learner is forced to divide visual attention between the assembly workstation itself, and a physically separated, often fine-print step-by-step instructions manual located elsewhere. A lot of visual scanning is involved and coupled sometimes with high cognitive demands in relating how a physical component, represented from the fixed viewpoint on the instructions, relates to the visual view of that component in the real assembly scene. This is a clear source of Extraneous Load (Mayer & Mareno, 2003) and hence a hindrance to learning. Instead, augmented reality on an HMD can be exploited by positioning the appropriate instruction in world (virtual) space, directly adjacent to the component to which it pertains; such close proximity between two items of information that need to be related to each other provides a compelling example of adherence to the proximity compatibility principle in systems design (Wickens & Carswell, 1995).

  In the current experiment, we employ virtual reality to depict the workspace for product assembly, rather than a physical system itself, because of the greater control this provides and because this mitigates some challenges of registering instructions onto a real physical device. Across two experiments, two groups of participants were taught to assemble eight different generic objects, which we refer to as "shapes", from a set of standard bars. Each shape contains 1 base bar (already placed) plus 5 consecutive bars that must be assembled in a proper sequence. Each bar in turn has either two or three attributes: Its length and the attachment point to connect to the previous bar (2 attributes), or the above two, coupled with a distinct color for each bar (3 attributes). This is designed to manipulate the Intrinsic Load of the shape construction. Extraneous Load is manipulated by the wordiness and the position of the instructions, with more straightforward instructions in the low condition.

*Germane Load* is manipulated, between two experiments, by whether an adaptive training regime is employed, to *optimize* Germane Load (Experiment 1), or a fixed difficulty increase (one size fits all) is employed, to create a non-optimal increase (Experiment 2). In both cases, the difficulty is increased by removing scaffolding. That is, on successive steps the printed instructions viewed on the screen are replaced by a requirement to remember the instruction, hence imposing the added cognitive (germane) load of working memory and rehearsal regarding which part must be connected where, to the previous part. In the adaptive training condition, if an error is made on an assembly, the participant is required to repeat that step, again consulting the initial instructions. If the step is performed correctly, assembly continues to the next step, and the previous one must be committed to memory for the next and for all future steps. Thus, the increase in Germane Load (requirement to remember the step before executing it), is contingent upon (adaptive to) each individual learner's proficiency. Hence, we label it "optimal" insofar as seeking each learner's level of desirable difficulty at each step. In the fixed difficulty increase (FDI) condition, we imposed a yoking design, by which the particular schedule of difficulty increase obtained by each participant in the adaptive condition, is stored, and then replayed to a yoked counterpart in the FDI condition. In this way, although not all participants in the FDI condition will receive the same schedule as all others, the difficulty increase obtained by all will not be adaptive to their performance and hence will be considered non-optimal.

**Experiment 1**

Experiment 1 aims to evaluate the effects of increased Extraneous and Intrinsic Load on workload and learning retention during the Assembly task, when participants are trained with optimal Germane Load by using performance-based adaptive training.

## Methods

**Participants.**

60 university students, aged between 18 and 56 ($M = 24.5$), were recruited from a university email list and compensated with a $30 Amazon gift card for their participation. Approximately 39 participants reported studying Computer Science as their major, with the rest being spread throughout various university programs. Forty-nine reported having used a VR headset before at least once. This research complied with the American Psychological Association Code of Ethics and was approved by the Institutional Review Board at Colorado State University. Informed consent was obtained from each participant.

**Study Design.**

This experiment utilized a 2 (Intrinsic Load: Low vs. High) by 2 (Extraneous Load: Low vs. High) within-subjects design to investigate how increasing either Intrinsic or Extraneous Load would impact overall training time, perceived mental workload, and retention of learned content. The four conditions resulting from the 2 × 2 design were counterbalanced between each of the eight-shapes, with two shapes corresponding to each of the four conditions.

**Independent Variables:**

*Intrinsic Load*. Manipulated by the number of attributes or "rules" that must be remembered when placing a bar to build the shape. See Figure 2 for a visual of each condition.

- <u>Low Intrinsic Load</u> (2-attributes): bar length and attachment point
- <u>High Intrinsic Load</u> (3-attributes): bar length, attachment point, and color.

*Extraneous Load*. Manipulated by the wordiness of step-by-step instructions, as well as the position of the instruction panel.

- <u>Low Extraneous Load:</u> "Take a medium red bar hole 3 and place it on the base bar hole 3."
- <u>High Extraneous Load:</u> "You will need to find a bar that is medium in size and red in color and then count to the third hole along the bar. The next step is to align the third hole of the bar with the third hole of the base bar and then join the two bars together in that spot." The instruction panel is offset to the left of the building area, as well as rotated 90 degrees, so that the participants will have to actively rotate away from the work, to read the instructions.

**Dependent Variables:**

*Mental Workload* was assessed with the NASA-TLX (Task Load Index; Hart & Staveland, 1988) Mental Workload subscale, on which participants rate their perceived demand associated with learning to build each shape on a scale from zero to twenty. The TLX is a widely used, multidimensional tool designed to assess subjective workload experienced by individuals during task performance. Previous studies have used direct europhysiological measures such as functional near-infrared spectroscopy (fNIRS) or electroencephalography (EEG) to measure workload and engagement (Hirshfield et al., 2023, and McDonnell et al., 2023).

For the current experiment, we chose to employ a self-report measure rather than a physiological measure to investigate whether the participants were in tune with the changes in workload imposed on them. As self-report measures are more easily deployed in operational settings than physiological measures, this also provides us with insight as to how we may utilize self-rated workload to facilitate optimal Germane Load in future studies and field settings.

**Procedure**

Figure 4 provides a diagram of the experimental procedure. Participants began the experiment by completing a demographics form and then entering the virtual environment through the HMD, where they received an auditory explanation of the task controls, procedure, and requirements. To ensure proficiency with the controls, participants proceeded to the practice task, where they constructed two shapes (with Intrinsic Load manipulation) with the researcher's help. After the practice task, participants were given the opportunity to ask clarifying questions and the experimenter reiterated important aspects of the task. Users then proceeded to the experimental task, where they were to learn to assemble eight unique shapes as outlined in Figure 3. For each of the eight shapes, participants were instructed on the assembly sequence over six cycles. Within each of these six cycles, participants received step-by-step instructions on which bar was placed next in the sequence. As the cycles progressed, another instruction was removed, requiring the participant to rely on their memory for the next step in the sequence. If the participant placed a bar incorrectly, they would be shown a red "X" and the instructions would reappear. After completing the six cycles of training, participants were given a 90-second distractor task of counting the number of times a ball bounces on the screen before taking the retention test for that shape. The retention test required participants to build the shape once without any instructions or feedback. The experimenter would then verbally administer the NASA-TLX Mental Demand scale. This process is repeated for each of the 8 to-be-learned shapes.

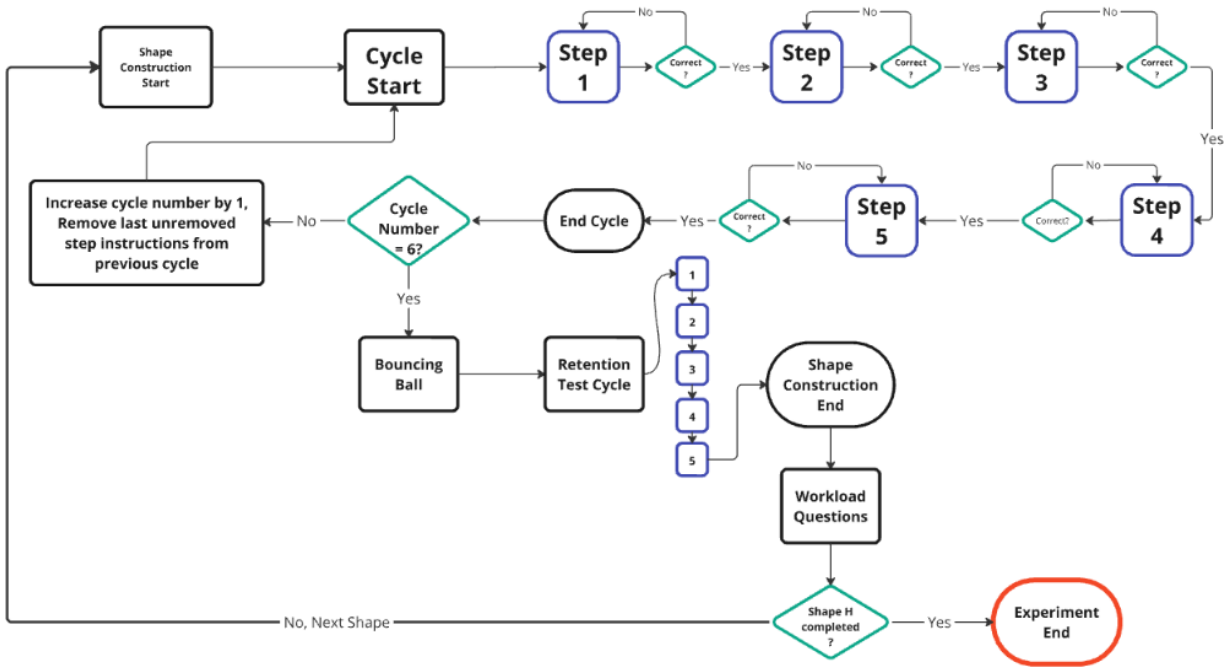

*Figure 3*. Experimental 1 Procedure.

# Results

**Training Time.** A 2 × 2 repeated measures ANOVA examining the effects of Intrinsic and Extraneous Load on training time revealed a significant main effect of Intrinsic Load ($F(1,58) = 53.74$, $p < .001$, $\eta^2 = .11$) and a significant main effect of Extraneous Load ($F(1,58) = 23.38$, $p < .001$, $\eta^2 = .05$). Increasing either type of load led to longer time spent training, therefore reducing training efficiency. No significant interaction was observed between the two types of loads ($F(1,58) = 1.13$, $p = .72$, $\eta^2 = .00$). Group means are reported in Figure 4.

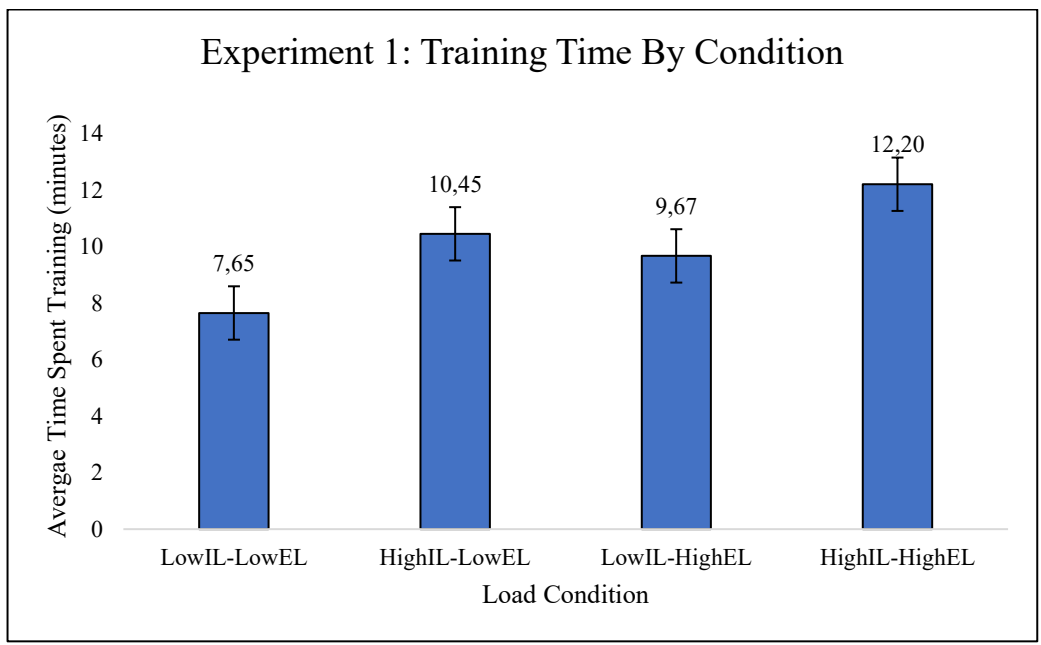

*Figure 4.* Experiment 1 Training time split by condition. Error bars represent the standard error of the mean.

*Note*: Intrinsic Load is abbreviated to IL and Extraneous Load is abbreviated to EL.

**Training Error Rate**. A 2 × 2 repeated measures ANOVA examining the effects of Intrinsic and Extraneous Load on error rate during training revealed a significant increase in error rate from 13.5 to 16.5 as Intrinsic Load increased ($F(1,58) = 9.22$, $p < .01$, $\eta^2 = .02$), but no significant main effect of Extraneous Load ($F(1,58) = 0.79$, $p = .38$, $\eta^2 = .003$). No significant interaction was observed between the two types of loads ($F(1,58) = 0.16$, $p = .69$, $\eta^2 = .00$). Group means are reported in Figure 5.

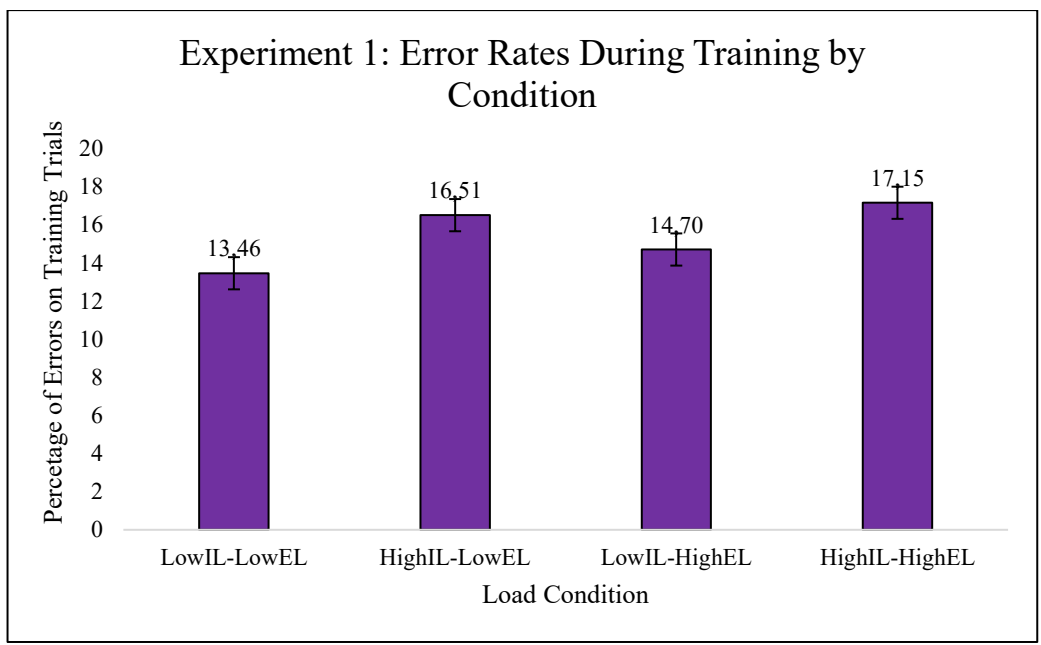

*Figure 5.* Experiment 1 Error rates during training by condition. Larger numbers indicate more errors made during training. Error bars represent the standard error of the mean.

*Note:* Intrinsic Load is abbreviated to IL and Extraneous Load is abbreviated to EL.

**Mental Workload.** A 2 × 2 repeated measures ANOVA revealed a significant main effect of Intrinsic Load ($F(1,58) = 13.05$, $p < .001$, $\eta^2 = .02$), but no significant main effect of Extraneous

Load ($F(1,58) = 0.46$, $p = .49$, $\eta^2 = .001$). No significant interaction was observed between the two types of loads ($F(1,58) = 0.60$, $p = .44$, $\eta^2 = .001$). Group means are reported in Figure 6.

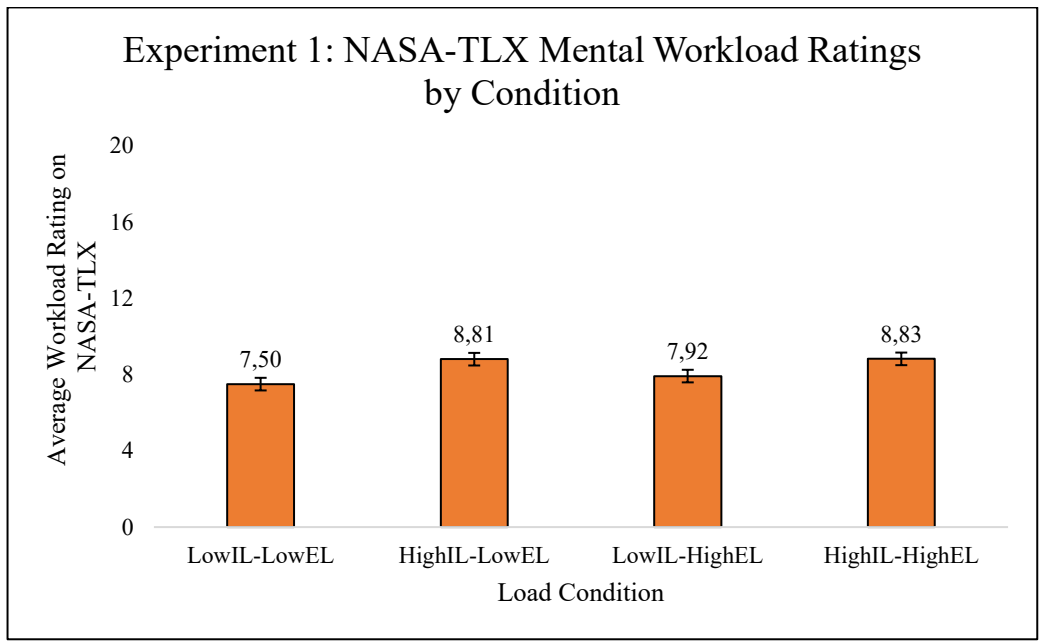

*Figure 6*. Experiment 1 NASA-TLX Mental Workload ratings by condition. Error bars represent the standard error of the mean.

*Note*: Intrinsic Load is abbreviated to IL and Extraneous Load is abbreviated to EL.

**Retention Test Time**. A 2 × 2 repeated measures ANOVA revealed no main effect of Intrinsic Load ($F(1,58) = 2.90$, $p = .09$, $\eta^2 = .01$) but did reveal a significant main effect of Extraneous Load ($F(1,58) = 42.17$, $p < .001$, $\eta^2 = .10$), increasing the time of the retention trial from 62 to 84 sec, an increase of 22 seconds.. No significant interaction was observed between the two types of loads ($F(1,58) = 1.70$, $p = .20$, $\eta^2 = .003$). Group means are reported in Figure 7.

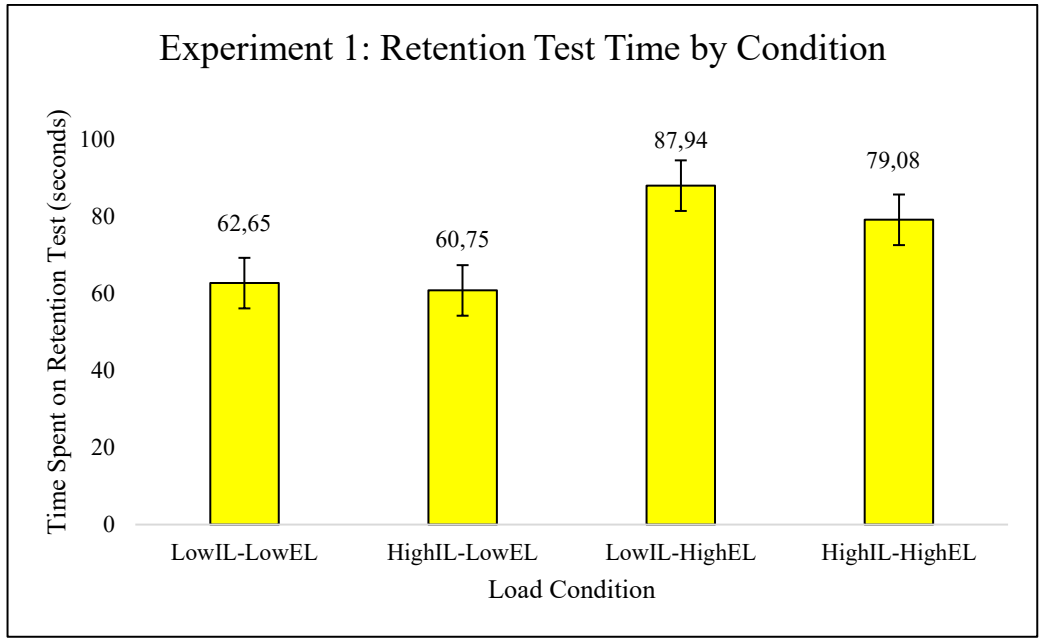

*Figure 7.* Experiment 1 retention test time split by condition. Error bars represent the standard error of the mean.

*Note:* Intrinsic Load is abbreviated to IL and Extraneous Load is abbreviated to EL.

**Retention Test Error Rate.** A 2 × 2 repeated measures ANOVA examining the effects of Intrinsic and Extraneous Load on retention test error rate revealed no main effect of Intrinsic Load ($F(1,58) = .012$, $p = .91$, $\eta^2 = .00$), nor Extraneous Load ($F(1,58) = 0.34$, $p < .56$, $\eta^2 = .001$). No significant interaction was observed between the two types of loads ($F(1,58) = 1.89$, $p = .17$, $\eta^2 = .01$). Group means are reported in Figure 8.

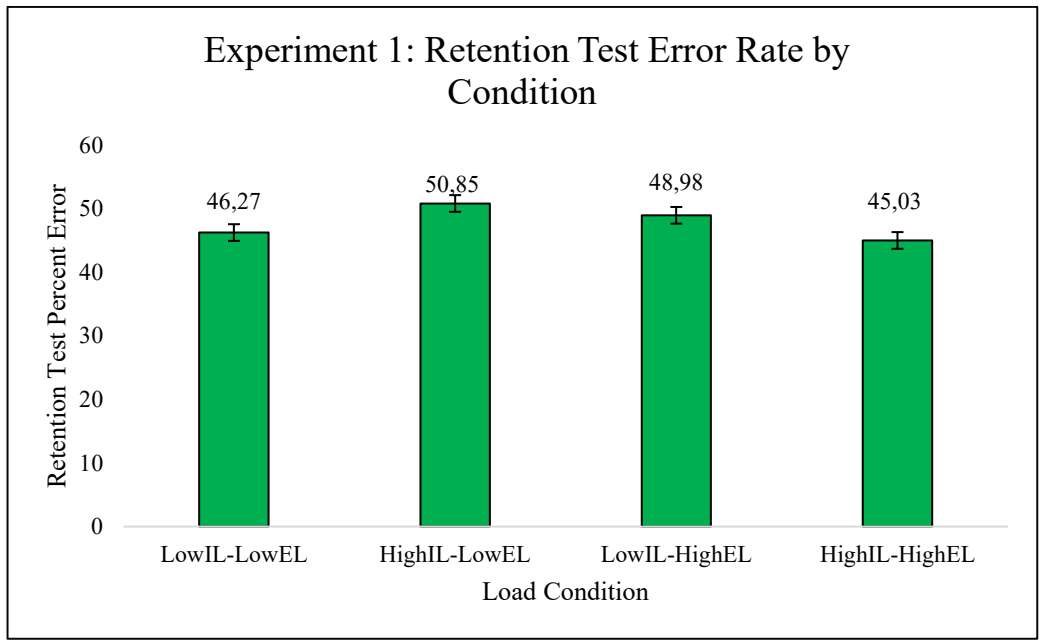

*Figure 8*. Experiment 1 Error rates on Retention Test by condition. Larger numbers indicate worse retention. Error bars represent the standard error of the mean.

*Note:* Intrinsic Load is abbreviated to IL and Extraneous Load is abbreviated to EL.

Tables 1 and 2 report the analyses of the impact of increasing each kind of load when the other type of load was low – not imposed. This analysis revealed that increasing Intrinsic Load increased training time, errors during training, and experienced mental workload. Increasing Extraneous Load also increased training time, but not errors during training or mental workload. The higher Extraneous Load did however have a detrimental effect on the time required to complete the retention trial.

**Table 1: Planned Contrasts examining the effects of increasing Intrinsic Load when Extraneous Load is low**

|  | Low | High | statistics |
|---|---|---|---|
| **Training time (minutes)** | 7.65 (*SD* = 2.93) | 10.45 (*SD* = 3.87) | $t(59) = -5.73, p < .001$*** |
| **Training Error Rate** | 13.50 (*SD* = 8.99) | 16.5 (*SD* = 14.3) | $t(59) = -2.58, p < .05$* |
| **TLX Workload** | 7.50 (*SD* = 4.03) | 8.82 (*SD* = 3.88) | $t(59) = -3.38, p < .001$*** |
| **Retention Test Time (seconds)** | 62.70 (*SD* = 30.20) | 60.8 (*SD* = 26.60) | $t(59) = 0.53, p = .60$ |
| **Retention Test Error Rate** | 46.3 (*SD* = 23.80) | 49.0 (*SD* = 22.50) | $t(59) = -1.17, p = .25$ |

**Table 2: Planned Contrasts examining the effects of increasing Extraneous Load when Intrinsic Load is low.**

|  | Low | High | statistics |
|---|---|---|---|
| **Training time (minutes)** | 7.65 (*SD* = 2.93) | 9.67 (*SD* = 3.83) | $t(59) = -4.65, p < .001$*** |
| **Training Error Rate** | 13.50 (*SD* = 8.99) | 14.70 (*SD* = 8.92) | $t(59) = -1.01, p = .32$ |
| **TLX workload** | 7.50 (*SD* = 4.03) | 7.92 (*SD* = 3.72) | $t(59) = -1.05, p = 0.30$ |
| **Retention Test Time (seconds)** | 62.70 (*SD* = 30.20) | 87.90 (*SD* = 37.70) | $t(59) = -6.02, p < .001$*** |
| **Retention Test Error Rate** | 46.3 (*SD* = 23.80) | 50.8 (SD = 25.70) | $t(59) = -0.67, p = .50$ |

## Discussion

The result of Experiment 1 was that increasing Intrinsic Load, through the working memory demands of the task, caused participants to experience a higher mental load (TLX), and this experience reduced the rate of skill acquisition significantly and substantially (by around 50%). However, the experience of this higher load did not seemingly disrupt the acquisition of the assembly skills as assessed by the learning retention test accuracy, contrary to predictions of CLT. This pattern of effects for Intrinsic Load was only partially replicated by the increase in Extraneous Load in that the TLX measure of subjective workload did not significantly increase with the wordier instructions of the higher load; and the increase in training time, while significant, was smaller than with the increase in Intrinsic Load. Like Intrinsic Load, the increase in Extraneous Load did not degrade performance accuracy on the retention test, but greater Extraneous Load in training did influence latency.

**Experiment 2**

Experiment 2 examined differences in learning outcomes when the imposed Germane Load was non-optimal versus the Experiment 1 situation when imposed Germane Load was tied to the individual learner (via adaptive training), and therefore, via achieving desirable difficulties, presumed to be optimal. Additionally, this experiment sought to investigate how the effects of Intrinsic and Extraneous Load differ based on variations in Germane Load. To accomplish this, Experiment 2 utilizes the same design as Experiment 1, however employing a fixed training system that does not reintroduce scaffolding in real time based on that learner's performance. Instead, the scaffolding removal/reintroduction schedule for each participant in Experiment 1 was yoked to a participant in Experiment 2. This yoking procedure is described in more detail below.

# Methods

**Participants.**

58 university students or employees, aged between 20 and 60 ($M = 27.1$), were recruited from a university email list and compensated with a $30 Amazon gift card for their participation. 22 participants reported majoring in Computer Science, and 41 reported having used a VR headset before.

**Study Design.**

Experiment 2 utilized the same 2 (Intrinsic Load: Low vs. High) by 2 (Extraneous Load: Low vs. High) within-subjects design as Experiment 1.

*Yoking.* In learning research, a yoked design is an experimental setup where two or more subjects are matched or "yoked" to ensure that they experience the same external conditions aside from the main experimental manipulation. In this experiment, yoking was employed to control the number of presentations of instructions participants received. This ensured that differences in learning outcomes could be attributed to the manipulation of Germane Load, and not simply that those who received adaptive training were given more or fewer opportunities to study the assembly instructions.

**Procedure.**

Participants in Experiment 2 followed a near-identical procedure to that in Experiment 1, but used a fixed, non-adaptive training system. Participants would receive tailored feedback on the accuracy of each bar placement, but only have scaffolding reintroduced and an opportunity to correct their mistake if their yoked counterpart had made a mistake at this point and had scaffolding returned for that step.

## Results

**Training Time.** A 2 × 2 repeated measures ANOVA examining the effects of Intrinsic and Extraneous Load on training time revealed a significant lengthening of training time imposed by increasing Intrinsic Load ($F(1,57) = 38.14, p < .001, \eta^2 = .12$) and a significant detrimental effect of increasing Extraneous Load ($F(1,57) = 21.15, p < .001, \eta^2 = .05$). No significant interaction was observed between the two types of loads ($F(1,57) = 0.67, p = .42, \eta^2 = .001$). Group means are reported in Figure 9.

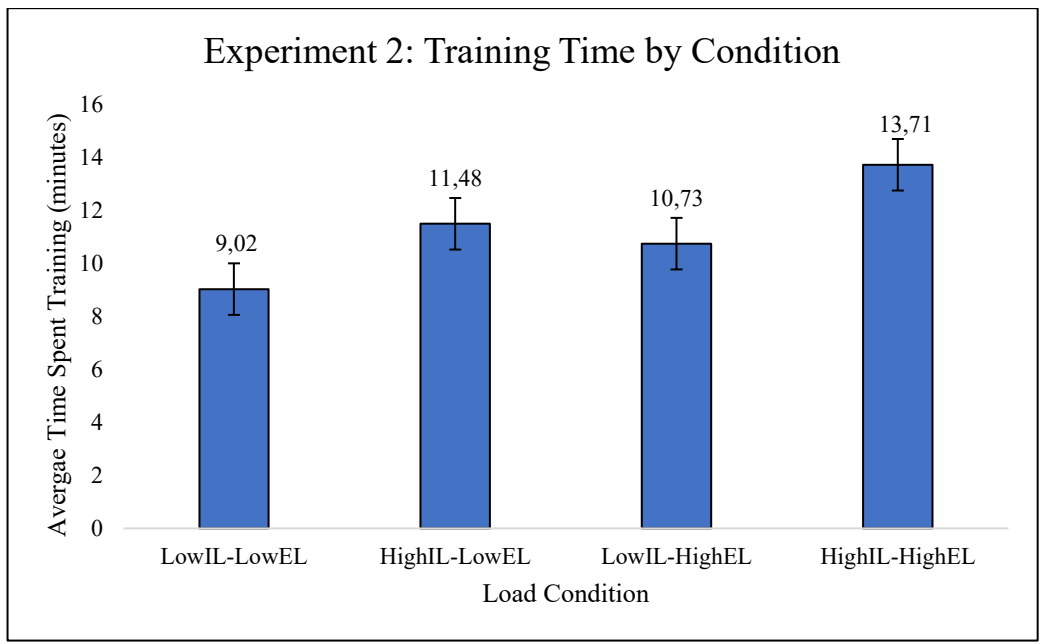

*Figure 9.* Experiment 2 Training time split by condition. Error bars represent the standard error of the mean.

*Note*: Intrinsic Load is abbreviated to IL and Extraneous Load is abbreviated to EL.

**Training Error Rate**. A 2 × 2 repeated measures ANOVA examining the effects of intrinsic and Extraneous Load on error rate during training revealed a significant increase in training errors resulting from the increase of Intrinsic Load ($F(1,57) = 17.68, p < .001, \eta^2 = .03$), but not from the increase in Extraneous Load ($F(1,57) = 0.11, p = .86, \eta^2 = .00$). No significant interaction was observed between the two types of loads ($F(1,57) = 1.87, p = .18, \eta^2 = .00$). Group means are reported in Figure 10.

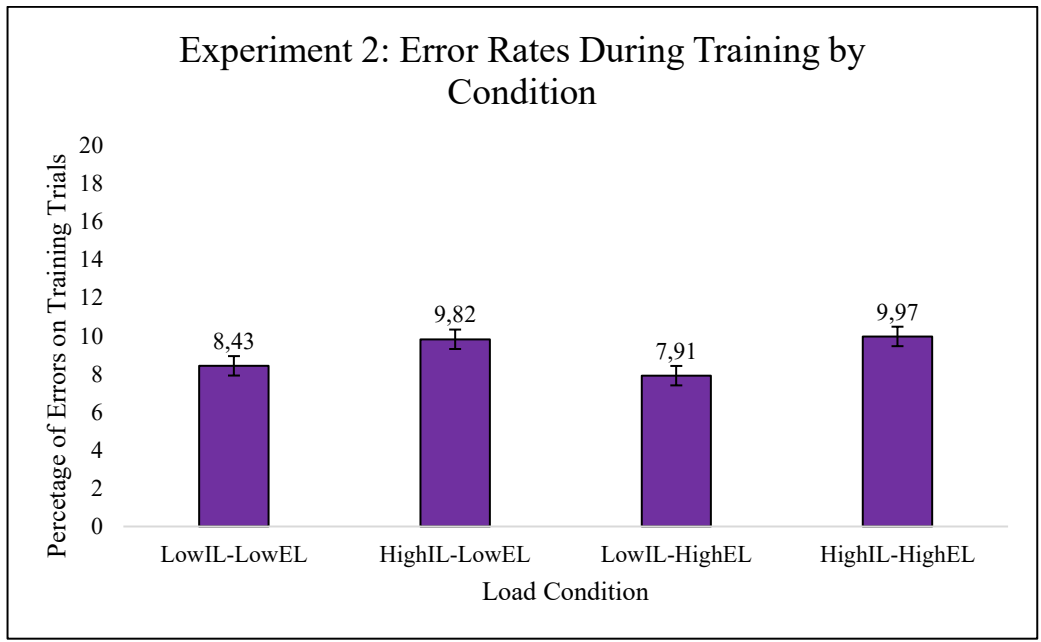

*Figure 10.* Experiment 2 error rates during training by condition. Larger numbers indicate more errors made during training. Error bars represent the standard error of the mean.

*Note*: Intrinsic Load is abbreviated to IL and Extraneous Load is abbreviated to EL.

**Mental Workload.** A 2 × 2 repeated measures ANOVA examining the effects of intrinsic and Extraneous Load on participants' subjective mental workload ratings on the NASA-TLX. This revealed a significant main effect of Intrinsic Load ($F(1,57) = 16.82$, $p < .001$, $\eta^2 = .03$) and a significant main effect of Extraneous Load ($F(1,57) = 7.89$, $p = < .01$, $\eta^2 = .01$). No significant interaction was observed between the two types of loads ($F(1,57) = 0.05$, $p = .83$, $\eta^2 = .00$). Group means are reported in Figure 11.

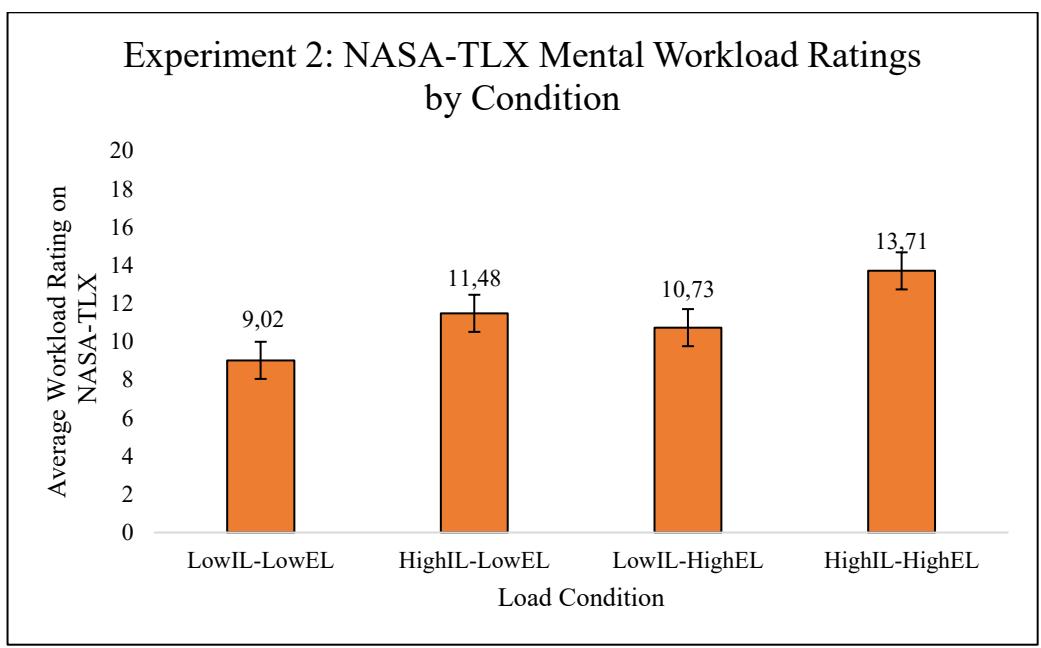

*Figure 11.* Experiment 2 NASA-TLX Mental Workload ratings by condition. Error bars represent the standard error of the mean.

*Note*: Intrinsic Load is abbreviated to IL and Extraneous Load is abbreviated to EL.

**Retention Test Time**. A 2 × 2 repeated measures ANOVA examining the effects of intrinsic and Extraneous Load on retention test response time revealed a significant main effect of Intrinsic

Load ($F(1,57) = 22.43$, $p = < .001$, $\eta^2 = .07$) but showed no effect of Extraneous Load ($F(1,57) = 0.04$, $p = .84$, $\eta^2 = .00$). No significant interaction was observed between the two types of loads ($F(1,57) = 0.002$, $p = .97$, $\eta^2 = .00$). Group means are reported in Figure 12.

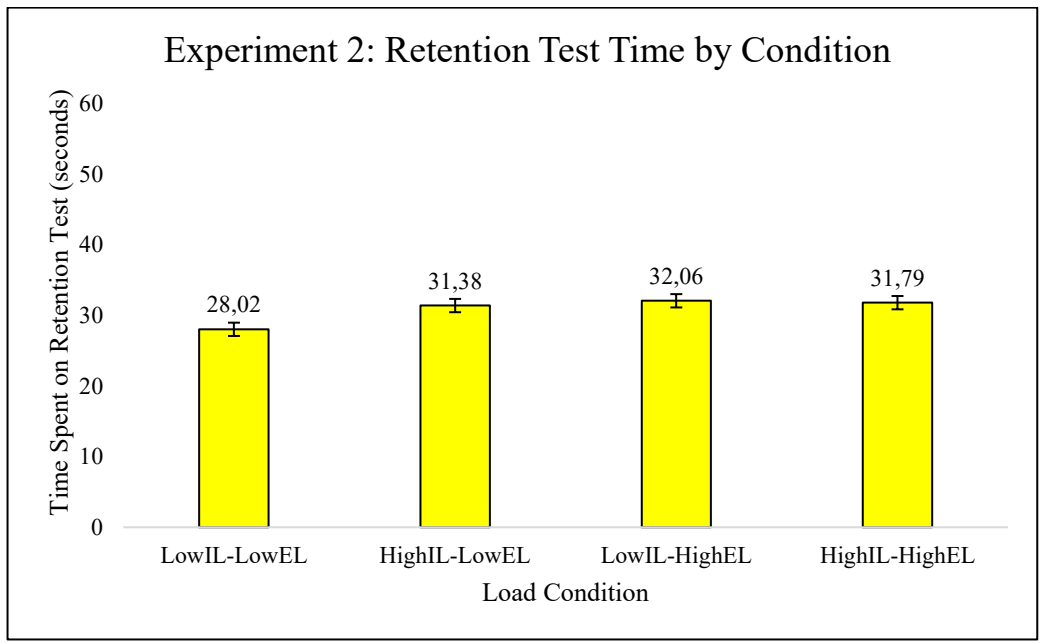

*Figure 12.* Experiment 2 retention test time split by condition. Error bars represent the standard error of the mean.

*Note*: Intrinsic Load is abbreviated to IL and Extraneous Load is abbreviated to EL.

**Retention Test Error Rate.** A 2 × 2 repeated measures ANOVA examining the effects of Intrinsic and Extraneous Load on retention error rate revealed no main effect of Intrinsic Load ($F(1,57) = 1.43$, p $= .24$, $\eta^2 = .01$) nor Extraneous Load ($F(1,57) = 1.75$, $p = .19$, $\eta^2 = .01$). No significant interaction was observed between the two types of loads ($F(1,57) = 0.66$, $p = .42$, $\eta^2 = .002$). Group means are reported in Figure 13.

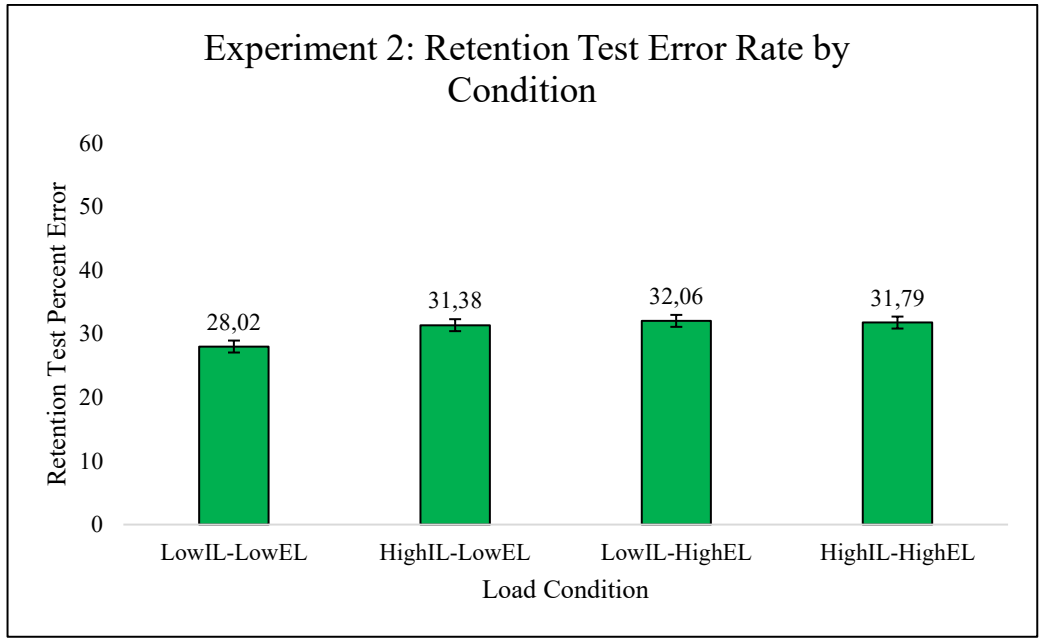

*Figure 13.* Experiment 2 Error rates on Retention Test by condition. Larger numbers indicate worse retention. Error bars represent the standard error of the mean.

*Note*: Intrinsic Load is abbreviated to IL and Extraneous Load is abbreviated to EL.

As with Experiment 1, planned contrasts were performed of the effect of increasing each kind of load, when the other source of load is not imposed, as shown in Tables 3 and 4.

**Table 3: Planned Contrasts examining the effects of increasing Intrinsic Load when Extraneous Load is low**

|  | Low | High | Statistics |
|---|---|---|---|
| **Training Time (minutes)** | 9.02 (*SD* = 2.90) | 11.48 (*SD* = 3.83) | $t(57) = -5.81, p < .001$*** |
| **Training Error Rate** | 8.43 (*SD* = 6.27) | 9.82 (*SD* = 6.34) | $t(57) = -2.31, p < .05$* |
| **TLX Workload** | 7.94 (*SD* = 3.57) | 9.20 (*SD* = 3.66) | $t(57) = -3.93, p < .001$*** |
| **Retention Test Time (seconds)** | 36.30 (*SD* = 15.50) | 45.90 (*SD* = 20.90) | $t(57) = -3.23, p < .01$** |
| Retention Test Error Rate | 28 (*SD* = 16.10) | 32.10 (*SD* = 18.60) | *t*(57) = -1.43, *p* = .16 |

**Table 4: Planned Contrasts examining the effects of increasing Extraneous Load when Intrinsic Load is low**

| | Low | High | Statistics |
|---|---|---|---|
| **Training Time (minutes)** | 9.02 (*SD* = 2.90) | 10.73 (*SD* = 3.24) | *t*(57) = -3.94, *p* <.001*** |
| **Training Error Rate** | 8.43 (*SD* = 6.27) | 7.91 (*SD* = 8.92) | *t*(57) = 0.73, *p* = .466 |
| **TLX Workload** | 7.94 (*SD* = 3.57) | 8.58 (*SD* = 3.79) | *t*(57) = -1.84, *p* = .07 |
| **Retention Test Time (seconds)** | 36.30 (*SD* = 15.50) | 87.90 (*SD* = 37.70) | *t*(57) = 0.36, *p* = .72 |
| **Retention Test Error Rate** | 28 (*SD* = 16.10) | 31.40 (*SD* = 20.50) | *t*(57) = -1.21, *p* = .23 |

**Germane Load Comparisons.** The primary purpose of Experiment 2 was to act as the comparison for the adaptive techniques of Experiment 1. We report below two analyses, the first is of this yoked design which now represents the 2 × 2 × 2 design of all three Independent Variables. Group means are reported in Table 5.

The 2 × 2 × 2 mixed measures ANOVA revealed no between-subjects effect of Germane Load Optimality (manipulated by training type) on either perceived workload ($F(1,115) = 1.07, p = .303, \eta^2 = .01$) or learning retention accuracy ($F(1,115) = 0.16, p = 0.69, \eta^2 = .01$.). This combined analysis of both experiments showed a main effect of Intrinsic Load, with increased both experienced mental workload ($F(1,115) = 29.82, p < .001, \eta^2 = .02$) and overall training time ($F(1,115) = 29.82, p < .001, \eta^2 = .02$); and the increase in Extraneous Load now *did* produce an increase in experienced workload ($F(1,115) = 4.88$, $p < .05$, $\eta2 = .003$), as well as an increase in training time ($F(1,115) = 4.38$, $p < .001$, $\eta2 = .06$).

The between-experiment analysis also revealed that implementing more optimal Germane Load via adaptive training (Experiment 1) shortened the average overall training time for all 8 shapes from 44.9 ($SD = 11.20$) minutes in the fixed-training condition to 39.6 ($SD = 10$) minutes in the adaptive training condition ($t(116) = - 2.58$, $p < .05$, $d = -0.48$).

Table 5. Group means included in the analysis.

|  |  | LowIL-LowEL | HighIL-LowEL | LowIL-HighEL | HighIL-HighEL |
|---|---|---|---|---|---|
| Mental Workload | Adaptive Training | 7.50 (SD = 4.03) | 8.81 (SD = 3.88) | 7.92 (SD = 3.72) | 8.82 (SD = 3.72) |
|  | Fixed Training | 7.94 (SD = 3.57) | 9.20 (SD = 3.66) | 8.58 (SD = 3.79) | 9.92 (SD = 4.41) |
| Retention Test Error Rate | Adaptive Training | 28.60 (SD = 14.90) | 30.90 (SD = 14.20) | 31.80 (SD = 16.10) | 28.10 (SD = 17.50) |
|  | Fixed Training | 32.30 (SD = 17.40) | 32.10 (SD = 18.60) | 31.40 (SD = 20.50) | 28 (SD = 16.10) |
| Training time (minutes) | Adaptive Training | 7.65 (SD = 2.93) | 10.40 (SD = 3.87) | 9.67 (SD = 3.83) | 12.20 (SD = 4.31) |
|  | Fixed Training | 9.02 (SD = 2.90) | 11.50 (SD = 2.85) | 10.70 (SD = 3.24) | 13.70 (SD = 5.06) |

The partial divergence of the effects of Extraneous Load increase between Experiment 1 (no effect on mental workload) and the combined analysis (significant, but weak, increase) reflects in part the greater statistical power of the combined analysis the finding of theoretical importance that increasing load in the non-optimal training environment of Experiment 2, is more noticeable than in the optimal training environment, achieved by adaptive training, in Experiment 1. This difference is seen in the much greater effect size of Intrinsic Load manipulations on increasing TLX in Experiment 2 ($\eta2 = .07$) than in Experiment 1 ($\eta2 = .02$) as well as the greater effect sizes of Extraneous Load manipulations on TLX in Experiment 2 ($\eta2$ = .03) than in Experiment 1 ($\eta2$ = .001).

## Discussion

The results of Experiment 2 demonstrate that Intrinsic Load influences training time, the rate of errors during training, and perceived mental workload in a fixed training environment. This pattern of results echoes Experiment 1, indicating that tasks with higher intrinsic cognitive demands affect learning efficiency and perceived difficulty but, in these cases, where extended practice time is available do not seem to impact actual learning retention (as evidenced by the lack of an effect on retention test error rates.) Similarly, Extraneous Load showed a significant impact on training time and mental workload, indicating that while it contributes to perceived task difficulty and slows acquisition time, it may not critically affect error rates or retention outcomes. This is, again, in contrast to the predictions of CLT. The effect of Extraneous Load on mental workload only being present when training is fixed and not tailored to the participants' current learning state may indicate that learners are more sensitive to distracting or unhelpful design choices in learning environments when they are not being provided adequate support from the system. The results of our between-experiment comparisons, speaking to the effect of optimization of Germane Load on learning are discussed below.

## General Discussion

The first purpose of the current set of two experiments was to examine the extent to which increasing the two types of loads, intrinsic and extraneous, would increase learner experienced load, increase training time and degrade the acquisition of knowledge as assessed by a retention test. We assessed experienced mental workload (TLX) as a manipulation check of that construct which is the foundation to cognitive load theory. We focused on training time because this measure is of relevance to any organization seeking to reduce training time without sacrificing learning. Finally, we measured retention test accuracy as this is a better index of knowledge acquisition than performance on a final learning trial (Soderstrom & Bjork, 2015). The second purpose of this experiment sequence was to examine how the effects of these two sources of load change when Germane Load is optimized to the learner through adaptive training versus when it is non-optimal in a fixed training scenario. We implemented greater optimization via adaptive training, employed to create "desirable difficulties" for each learner during the learning process (Bjork & Bjork, 2020). Such difficulty seeks to prevent any given learner from being either overwhelmed by a task that is too difficult, or from encountering a task so easy that there is little needed to invest in its learning. The linkage between expertise and CLT is well validated in research on the "expertise effect" of CLT (Paas & van Gog, 2009; Rey & Buchwald, 2011).

Regarding H1, the impact of intrinsic load on workload and learning performance was partially supported. Our results yielded success in producing a large increase of perceived load as inferred from the NASA TLX ratings. This is not surprising in that our manipulation directly imposed on working memory, well-validated to exert a profound influence on many measures of workload (Longo & Orrú, 2022, Hart & Wickens, 2010, Kahneman & Beatty, 1966). We assume therefore that participants had a difficult time mastering the memory of each shape when three attributes needed to be retained, an effect clearly revealed by their longer time spent training in the high intrinsic load conditions.

Contrary to our predictions in H1, increasing intrinsic load showed no performance cost to their retention of the assembly at test. This result may suggest that our participants were successful in compensatory learning behavior, a sort of speed-accuracy tradeoff in that they spent longer to master the more difficult High Intrinsic Load shapes than the Low Intrinsic Load shapes, in order to be able to obtain high accuracy. It is also possible that participants responded to the increased memory demands of three attributes as if this were a source of Germane Load.

Regarding H2, predicting similar effects of Extraneous Load by rendering the instructions with more verbosity, our results were again mixed, showing small and inconsistent effects across the two experiments in the between-subjects ANOVA. An increase in perceived workload was observed in the non-adaptive Experiment 2 but not observed in the adaptive training of Experiment 1. The fact that there was some, albeit muted, effect of Extraneous Load may be the result of participants using selective attention to rapidly filter through the instructions and find the key terms. This attentional strategy in turn proved to be very effective in Experiment 1, where training was more optimal. However, the resources saved by this strategy were not reflected in a less experienced workload but instead were redeployed to Germane Load to create more efficient learning. This deployment created a successful buffering of the Extraneous Load cost on the delayed test. There is also the possibility that participants used the occasional presence of visual representations of the objects they were building to augment their learning of layout of the parts. The possibility of studying the visual arrangement in this training context might have served to somewhat diminish the impact of the verbal instruction manipulations.

Our manipulation of Germane Load optimality did not yield all the expected or hypothesized effects in H3. Neither experiment found an effect of Germane Load on mental workload. This null effect might reflect that the resource availed by the more optimal training at the task's desirability difficult were not actually "saved" (which would have yielded a lower TLX score), but were instead redeployed to the learning process itself, thereby shortening its duration.

**Limitations and Future Directions**

The above discussion has explained the lack of an effect of any of the three loads on accuracy of delayed testing as a consequence of compensatory behavior on the part of the learner – spending longer if needed, to preserve long term learning and retention. However, it is also possible that our relatively short delay of 1 minute between final learning trial and test was of insufficient magnitude to fully demonstrate possible training benefits. In other examinations of training strategy benefit success, for example of the "testing effect" (Roediger & Karpicke, 2006) or the benefits of spaced over massed training (Dunlosky, et al., 2013), the delay between the final training session and the test is often considerably longer like a matter of days. Indeed, for the testing effect, Roediger and Karpicke (2006) found that the shortest delay indicated that re-reading produced better performance than self-testing. While our current strategy of imposing the 90-second ball counting "filler" task, was assumed to be sufficient to prevent rehearsal, and hence eliminate any residual content in working memory at the time of the test, future studies should examine how further consolidation of a learned skill could proceed over a much longer time period.

## Conclusion

These findings suggest that increasing Intrinsic Load reliably affects learner-perceived workload and skill acquisition time in the current assembly task context. The mixed effects of Extraneous Load underscore the role of selective attention, as learners appeared to effectively filter unnecessary information, particularly in adaptive training settings, thereby preserving efficiency. Finally, the limited influence of Germane Load optimization may reflect participants' ability to redirect cognitive resources toward learning rather than workload reduction, though future studies should explore longer retention intervals to capture delayed test effects fully. Perhaps most importantly, these experiments demonstrate that adaptive training systems can produce more efficient training times without cost to training outcomes.

## Key Points

- In this paradigm, participants learned to assemble 8 unique shapes in virtual reality with varying levels of Intrinsic Load (shape complexity), Extraneous Load (instruction verbosity), and Germane Load (adaptive vs. fixed training).
- Increasing intrinsic load heightened perceived workload and extended training time, but did not degrade retention test performance, suggesting compensatory learning strategies. Extraneous load effects were mixed, with selective attention helping mitigate its impact, particularly in adaptive training.

- Germane load optimization did not reduce subjective workload as hypothesized but did enhance training efficiency by shortening training duration without cost to learning retention.
- These results have implications for keeping instructional prompts concise and early interpretable instructions during training as well as utilizing adaptive techniques to improve training efficiency.

# Biographies

**Rebecca L. Pharmer** is a doctoral candidate at Colorado State University. She received her Master of Science in Psychology from Colorado State University in 2023.

**Christopher D. Wickens** is a professor emeritus of aviation and psychology at the University of Illinois and is a senior scientist at Alion Science and Technology, Boulder, Colorado, and a professor of psychology at Colorado State University. He received his PhD in Psychology from University of Michigan in 1974.

**Lucas Plabst** is a researcher at Colorado State University, USA. He has a Doctorate and PhD in Computer Science from Colorado State University and the University of Würzburg, Germany. He received his Master of Science degree in Human-Computer-Interaction from the University of Würzburg in 2021.

**Benjamin A. Clegg** is a professor of cognitive psychology at Montanna State University. He received his PhD in Psychology in 1998 from the University of Oregon.

**Leanne M. Hirshfield** is an associate research professor at the University of Colorado, Boulder. She received her PhD in computer science from Tufts University in 2009.

**Joanna I. Lewis** is an assistant professor at the University of Northern Colorado. She received her PhD in Cognitive and Human Factors Psychology from the University of Central Florida.

**Jaylnn B. Nicoly** is a first-year Ph.D. student at the University of Colorado Boulder advised by Leanne Hirshfield. She received her Bachelor's degree in Computer Science from Colorado State University in 2024.

**Cara A. Spencer i**s a doctoral student at the University of Colorado Boulder in Computer Science. She received her Bachelor's degrees in Cognitive Science and Human Developmental Sciences from University of California San Diego in 2019. Before coming to Boulder, she worked as a research associate at NASA's Behavioral Health and Performance Laboratory.

**Francisco R. Ortega** is an associate professor at Colorado State University in Computer Science and director of the Natural User Interaction Lab (NUILAB). He earned his PhD in computer science in human computer interaction and 3D user interfaces from Florida International University (FIU). He also held the positions of postdoc and visiting assistant professor at FIU from February 2015 to July 2018.